\documentclass[10pt]{iopart}
\usepackage{epsf}

\begin{document}

\title[Little-Parks effect in hybrid superconductor--ferromagnet system]
{Little-Parks effect and multiquanta vortices in a hybrid
superconductor--ferromagnet system}

\author{A Yu Aladyshkin, A S Mel'nikov, D A Ryzhov}
\address{Institute for Physics of Microstructures,
 Russian Academy of Sciences, 603950, Nizhny Novgorod, GSP-105, Russia}
\ead{alay@ipm.sci-nnov.ru}

\begin{abstract}
Within the phenomenological Ginzburg-Landau theory we investigate the
phase diagram of a thin superconducting film with ferromagnetic
nanoparticles. We study the oscillatory dependence of the critical
temperature on an external magnetic field similar to the Little-Parks
effect and formation of multiquantum vortex structures. The structure of a
superconducting state is studied both analytically and numerically.
\end{abstract}

\submitto{\JPCM}
\maketitle

\nosections

Little-Parks effect \cite{Little-Parks}, i.e. oscillations of the critical
temperature $T_{\rm c}$ of multiply--connected superconducting samples in an
applied magnetic field $H$, is one of the striking phenomenon
demonstrating coherent nature of superconducting state. Such oscillatory
behaviour of $T_{\rm c}(H)$ is not a distinctive feature of a superconducting
thin--wall cylinder and can be observed also in superconductors with
columnar defects and holes (see references \cite{Buzdin,Bezr1,Bezr2}) and
mesoscopic simply--connected samples of the size of several coherence
lengths \cite{Meso,Chibotaru}. Generally, the oscillations of $T_{\rm c}$ with a
change in the external magnetic flux are caused by the transitions between
the states with different vorticities (winding numbers) characterizing the
circulation of the phase of the order parameter. For a system with
cylindrical symmetry the vorticity parameter just coincides with the
angular momentum of the Cooper pair wave function. The states with a
certain angular momentum $m$ can be considered as $m$-quanta vortices.
Experimental and theoretical investigations of these exotic vortex
structures (multiquanta vortices and vortex molecules) in mesoscopic
superconductors have attracted a great deal of attention. As we change an
external homogeneous magnetic field multiquanta vortices and vortex
molecules can transform one into another via first or second order phase
transitions.

In this paper we focus on another possibility to create multiquantum
vortex states: nucleation of superconducting order parameter in a hybrid
system consisting of a thin superconducting film and an array of magnetic
nanoparticles. The interest to such structures is stimulated by their
large potential for applications (e.g., as switches or systems with a
controlled artificial pinning). The enhancement of the depinning critical
current density $j_{\rm c}$ has been observed experimentally for superconducting
films with arrays of submicron magnetic dots \cite{Martin,Morgan,Bael-2},
antidots \cite{Bael-1}, and for superconductor--ferromagnet (S/F) bilayers
with domain structure in ferromagnetic films \cite{Santiago}. The matching
effects observed for magnetic and transport characteristics were explained
in terms of commensurability between the flux lattice and the lattice of
magnetic particles. Vortex structures and pinning in the S/F systems at
rather low magnetic fields (in the London approximation) have been
analysed in papers
\cite{sonin,iosif,Lyuksyutov,Sasik,Bulaevsky,Bespyatykh,Erdin,Helseth,Milosevic-2,Sonin-2}.

Provided the thickness of a superconducting film is rather small as
compared with the coherence length, the critical temperature of the
superconducting transition as well as the structure of superconducting
nuclei should be determined by a two--dimensional distribution of a
magnetic field component $B_{\rm z}(x,y)$ (perpendicular to the
superconducting film plane) induced by the ferromagnetic particles.
Obviously, the highest critical temperature corresponds to the nuclei
which appear near the lines of zeros of $B_{\rm z}$ due to the mechanism
analogous to the one responsible for the surface superconductivity (see,
e.g., reference \cite{sj}) and domain wall superconductivity
\cite{buzdin,buzdin2,un}. Provided these lines of zeros have the shape of
closed loops, the winding number of a superconducting nucleus will be
determined by the magnetic flux through the loop. Thus, changing this flux
(e.g., increasing an external $H$ field applied along the $z$ axis) we can
control the winding number. The resulting phase transitions between the
multiquantum states with different $m$ can cause the oscillations of
$T_{\rm c}$. Such oscillatory behaviour has been, in fact, observed in
reference \cite{Otani} for a Nb film with an array of GdCo particles. Note
that a change in a slope of the phase transition curve $T_{\rm c}(H)$
(which is probably a signature of the transition discussed above) has been
also found in reference \cite{Lange} for a Pb film with CoPd particles.
Provided the dimensions of the sample in the $(xy)$ plane are compared
with the coherence length, we can expect a rather complicated picture
which is influenced both by the sample edges and by the distribution of an
inhomogeneous magnetic field. For a model step-like profile of the
magnetic field the resulting phase transitions between different types of
exotic vortex states in a mesoscopic disc have been studied numerically in
reference \cite{Milosevic}. The interplay between the boundary effects and
magnetic field inhomogeneity influences also the formation of multiquantum
vortex states around a finite size magnetic dot embedded in a large area
superconducting film \cite{cheng,marm}. The transitions between different
multiquantum vortex states with a change in magnetic field and magnetic
dot parameters were studied in references \cite{Milosevic,cheng,marm} for
certain temperature values. These effects are closely related to the ones
observed in mesoscopic and multiply-connected samples, and consequently,
we can expect that the oscillations of $T_{\rm c}(H)$ (analysed below)
should also be a common feature of multiply-connected superconductors and
thin-film systems with magnetic dots.

In this work we do not consider the magnetic phase transitions in the
mixed state for $T<T_{\rm c}$ and focus on the oscillatory behaviour of $T_{\rm c}(H)$
in a large area superconducting film caused only by the quantization
associated with the characteristics of the inhomogeneous magnetic field
produced by ferromagnetic particles. We neglect the influence of the edge
and proximity effects in the S/F system and consider a nanoparticle only
as a source of a small-scale magnetic field. Our further consideration is
based on the linearized Ginzburg-Landau model:
\begin{equation}
 \label{GL-eq}
 - \left(\nabla+\frac{2\pi {\rm i}}{\Phi_0}{\bf A}\right)^2\Psi
 = \frac{1}{\xi^2(T)}\Psi \ .
\end{equation}
Here $\Psi({\bf r})$ is the order parameter, ${\bf A}({\bf r})$ is the
vector potential, ${\bf B}({\bf r})=\nabla\times{\bf A}({\bf r})$,
$\Phi_0$ is the magnetic flux quantum, $\xi(T)=\xi_0/\sqrt{1-T/T_{\rm
c0}}$ is the coherence length, and $T_{\rm c0}$ is the critical
temperature of the bulk superconductor at $B=0$. For the sake of
simplicity we neglect the interference effects between the superconducting
nuclei appearing near different nanoparticles (i.e. assume the
interparticle distance to be rather large as compared with the
superconducting nucleus size) and consider a single magnetic particle with
a fixed magnetic moment chosen perpendicular to the film plane $xy$. For a
rather thin film (of the thickness less than the coherence length) we can
neglect the influence of the field components $B_{\rm x},B_{\rm y}$ in the
film plane and consider an axially symmetrical two--dimensional problem
(\ref{GL-eq}) in the field $B_{\rm z}(r)=H+b(r)$, where $b(r)$ is the
$z$--component of the field induced by the ferromagnetic particle and
$(r,\theta,z)$ is a cylindrical coordinate system. Choosing the gauge
$A_{\theta}(r)=Hr/2+a(r)$ one can find the solution of the equation
(\ref{GL-eq}) in the form $\Psi({\bf r})=g_{m}(r)\exp({\rm i}
m\theta)/\sqrt{r}$, where $m$ is the vorticity, and $g_m(r)\propto
r^{|m|+1/2}$ for $r\rightarrow0$. The function $g_{m}(r)$ should be
determined from the equation:
\begin{equation}
\label{Schr}
 -\frac{{\rm d}^2g_m}{{\rm d}r^2} + \left[
 \frac{(\Phi(r)/\Phi_0+m)^2}{r^2}-\frac{1}{4r^2}\right]g_m
 = \frac{1}{\xi^2(T)}g_m \ .
\end{equation}
Here $\Phi(r)=2\pi rA_\theta(r)$ is the total flux through the circle of
radius $r$. The lowest eigenvalue $1/\xi^2(T)$ of the Schr\"odinger-like
equation (\ref{Schr}) defines the critical temperature $T_{\rm c}$ of the
phase transition into a superconducting state.

Obviously, for rather small fields $H$ the superconducting order parameter
can nucleate either far from the magnetic particle ($r\rightarrow\infty$)
where the critical temperature $T_{\rm c}^{\rm H}$ is defined by the
homogeneous field $B_{\rm z}=H$ or in the region close to the circle of
the radius $r_0$ where $B_{\rm z}(r_0)=0$ and $T_{\rm c}$ is controlled by
the slope of $B_{\rm z}(r)$ at $r=r_0$ and by the flux through the area of
the radius $r_0$. In the first case we obtain $1-T_{\rm c}^{\rm H}/T_{\rm
c0}=2\pi |H|\xi_0^2/\Phi_0$. For the second case we can analyse the
behaviour of $T_{\rm c}(H)$ assuming that the characteristic length scale
$\ell$ of the order parameter nucleus is much less than the characteristic
scale of the magnetic field distribution. Within such local approximation
(similar to the one used in reference \cite{un} for the description of
domain wall superconductivity) we can expand the flux in powers of the
distance from $r_0$:
 $$
 \frac{\Phi(r)}{\Phi_0}+m
 \simeq \left(\frac{\Phi(r_0)}{\Phi_0}+m\right)
 + \frac{\pi r_0B_{\rm z}^\prime(r_0)}{\Phi_0}(r-r_0)^2 \ .
 $$
This local approximation is valid under the following conditions:
 $$
 \left|\frac{B_{\rm z}^{\prime\prime}(r_0)}{B_{\rm z}^\prime(r_0)}\ell\right| \ll 1
 \qquad{\rm and}\qquad
 \frac{\ell}{r_0} \ll 1.
 $$

Introducing a new coordinate $t=(r-r_0)/\ell$ we obtain the dimensionless
equation
\begin{equation}
 \label{Dim-less}
 -\frac{{\rm d}^2g}{{\rm d}t^2} + (t^2-Q)^2g = Eg \ ,
\end{equation}
where the parameters $E$ and $Q$ are given by the expressions:
\begin{equation}
\label{Energy}
 \fl
 E = \frac{\ell^2}{\xi_0^2}\left(1-\frac{T}{T_{\rm c0}}\right) \ ,\quad
 \ell = \sqrt[3]{\frac{\Phi_0}{\pi \left|B_{\rm z}^\prime(r_0)\right|}} \ ,
 \quad
 Q = -\left(\frac{\Phi(r_0)}{\Phi_0}+m\right)
 \sqrt[3]{\frac{\Phi_0}{\pi r_0^3 B_{\rm z}^\prime(r_0)}} \ .
\end{equation}

We obtain $E(Q)\simeq Q^2+\sqrt{-2Q}$ when $Q\ll -1$, and $E(Q)\simeq
2\sqrt{Q}$ when $Q\gg 1$. The minimal value of $E(Q)$ is $E=E_{\rm
min}\simeq 0.904$ at $Q\simeq0.437$. The final expression for the critical
temperature reads:

\begin{equation}
\label{ana}
 \fl
 1 - \frac{T_{\rm c}}{T_{\rm c0}} = \frac{\xi_0^2}{\ell^2}
 \left[\min_m E\left(-\left(\frac{\Phi(r_0)}{\Phi_0}+m\right)
 \sqrt[3]{\frac{\Phi_0}{\pi r_0^3 B_{\rm z}^\prime(r_0)}}\right)
 + O\left(\frac{\ell^2}{r_0^2}\right)
 \right] \ .
\end{equation}
The superconducting nuclei are localized near the ferromagnetic particle
at a distance $r_0$. The states with different energetically favorable
winding numbers $m$ correspond to the multiquantum vortex structures very
similar to the ones observed in a mesoscopic disc. As we change an
external field $H$, we change the flux $\Phi(r_0)$ and, thus, change the
energetically favorable vorticity number and position of the nucleus.

To investigate the details of the oscillatory behaviour discussed above we
consider a particular case of a small ferromagnetic particle which can be
described as a point magnetic dipole with a magnetic moment ${\bf M}=M{\bf
z}_0$ placed at a height $h$ over the superconducting film. The
corresponding expressions for the field and the vector potential are:
\begin{equation}
\label{Dipole}
 b(r) = \frac{M(2h^2-r^2)}{(r^2+h^2)^{5/2}} \ ,\quad
 a(r) = \frac{Mr}{(r^2+h^2)^{3/2}} \ .
\end{equation}

\begin{figure}[htb]
\begin{center}
\epsfxsize=130mm
\epsfbox{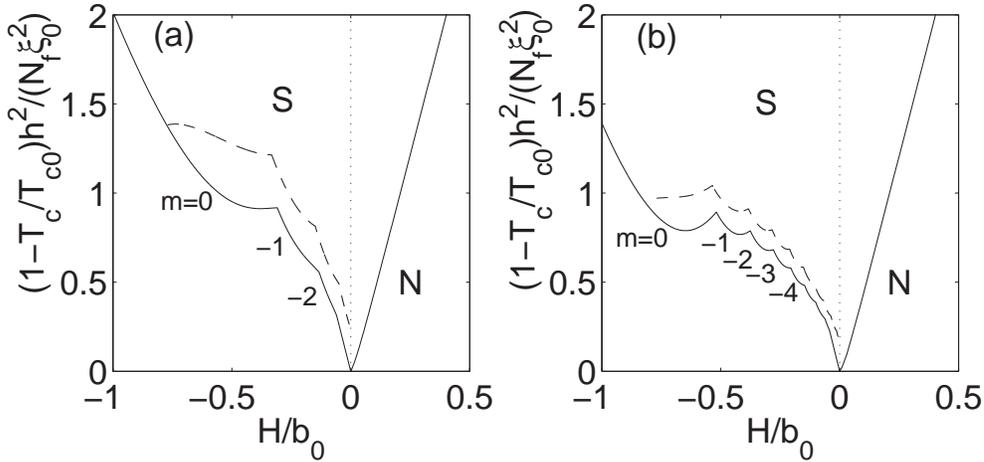}
\end{center}
\caption{
 \label{fig1ab}
Critical temperature as a function of an external magnetic field for
$N_{\rm f}=4$ (a) and $N_{\rm f}=10$ (b). Solid line is a result of direct
numerical simulations of equation (\ref{numer}). Dash line is obtained
from the analytical formula (\ref{ana}). Certain winding numbers $m$ for
different parts of the phase transition line are shown.}
\end{figure}

Introducing $f_m(r)=g_m(r)/\sqrt{r}$ and a dimensionless coordinate
$\rho=r/h$ we obtain the equation (\ref{Schr}) in the form:
\begin{equation}
\label{numer}
\fl
 - \frac{1}{\rho}\frac{\rm d}{{\rm d} \rho}
 \left(\rho\frac{{\rm d} f_m}{{\rm d} \rho}\right)
 + \left(\frac{3\sqrt{3}}{2}N_{\rm f}\left[
 \frac{H}{b_0}\rho +
 \frac{\rho}{(1+\rho^2)^{3/2}}\right]
 +\frac{m}{\rho}\right)^2 f_m =
 \frac{h^2}{\xi_0^2}\left(1-\frac{T_{\rm c}}{T_{\rm c0}}\right)f_{m} \ ,
\end{equation}
where $N_{\rm f}=4\pi M/(3\sqrt{3}h\Phi_0)$ is the dimensionless flux through
the area with the positive field $b(r)$ and $b_0=b(0)=2M/h^3$. In the
limit of small fields $H\rightarrow 0$ the nucleation of superconductivity
occurs at large distances $\rho$ and critical temperatures for different
winding numbers $m$ are very close. Thus, in this limit the critical
temperature is equal to $T_{\rm c}^{\rm H}$ and is not sensitive to the
presence of the dipole. Below this temperature we obtain a lattice of
singly quantized vortices (with the concentration determined by $H$) which
is surely disturbed under the dipole. Note, that the behaviour of $T_{\rm
c}$ in this low field regime should modify provided we take account of a
finite distance between the magnetic particles. For large absolute values
$|H|$ (much larger than the maximum field induced by the dipole) we obtain
the following asymptotical behaviour of $T_{\rm c}$: $1-T_{\rm c}/T_{\rm
c0}=2\pi\xi_0^2(-H-b_0)/\Phi_0$ for negative $H$ and $1-T_{\rm c}/T_{\rm
c0}=2\pi\xi_0^2(H-b_0/(25\sqrt{5}))/\Phi_0$ for positive $H$ values (here
$-b_0/(25\sqrt{5})$ is the minimum of the dipole field). The
superconductivity nucleates near the minima of the total field $|B_{\rm
z}|$ and, thus, is localized near the dipole. In the intermediate field
region ($-1<H/b_0<1/(25\sqrt{5})$) we should expect the oscillatory
behaviour of $T_{\rm c}$ discussed above. The number of oscillations is
controlled by the parameter $N_{\rm f}$. We have carried out the numerical
calculations of equation (\ref{numer}) for the various $N_{\rm f}$ values. For
the numerical analysis of the localized states of equation (\ref{numer})
we approximated it on a equidistant grid and obtained the eigenfunctions
$f_m(\rho)$ and eigenvalues by the diagonalization method of the
tridiagonal difference scheme. The results of these calculations as well
as the analytical dependence of $T_{\rm c}$ given by the expression
(\ref{ana}) are shown in figure \ref{fig1ab}.

We observed a remarkable asymmetry of the phase transition curve ($T_{\rm
c}(H)\neq T_{\rm c}(-H)$) which is caused by the difference in the
distributions of positive and negative parts of the dipole field $b(r)$:
the maximum positive field ($b_0$) is much larger than the absolute value
of the minimum negative field ($b_0/(25\sqrt{5})$). As a result, the
$T_{\rm c}$ oscillations appear to be most pronounced for negative $H$
which compensate the positive part of the dipole field. Taking $M\sim
3\times 10^{-11}\ G \cdot cm^3$ (for a ferromagnetic particle with
dimensions $300\ nm \times 300\ nm \times 300\ nm$ and magnetization $\sim
10^3\ G$), $h\sim 300\ nm$ we obtain $N_{\rm f}\simeq 10$, $b_0\sim 10^3 \
G$ and the characteristic scales of $T_{\rm c}$ oscillations $\Delta H\sim
100\ Oe$, $\Delta T_{\rm c}\sim 10^{-2}T_{\rm c0}\sim 0.1\ K$ for a Nb
film with $\xi_0\sim 40\ nm$ and $T_{\rm c0}\sim 8 \ K$.

We expect that the oscillatory behaviour of $T_{\rm c}$ can be observable,
e.g., in magnetoresistance measurements of thin superconducting films with
arrays of ferromagnetic particles. The superconducting nuclei localized
near the particles should result in the partial decrease in the resistance
below the oscillating $T_{\rm c}(H)$. As we decrease the temperature below
$T_{\rm c}(H)$ the supeconducting order parameter around a single particle
becomes a mixture of angular harmonics with different $m$ values and we
can expect the appearance the phase transitions similar to the ones
discussed in papers \cite{Milosevic,marm}. With the further decrease in
temperature the whole film becomes superconducting and resistivity turns
to zero.

\ack
We would like to thank A.~I.~Buzdin, A.~A.
Fraerman, Yu.~N. Nozdrin, and I.~A. Shereshevskii for stimulating
discussions. This work was supported, in part, by the Russian Foundation
for Basic Research, Grant No. 03-02-16774, Russian Academy of Sciences
under the Program 'Quantum Macrophysics', Russian State Fellowship for
young doctors of sciences (MD-141.2003.02), University of Nizhny Novgorod
under the program BRHE and 'Physics of Solid State Nanostructures'.

\end{document}